# Non-invasive thermal comfort perception based on subtleness magnification and deep learning for energy efficiency

Xiaogang Cheng[1, 4, 5], Bin Yang[2, 3, *], Anders Hedman[4], Thomas Olofsson[3], Haibo Li[1, 4], Luc Van Gool[5]

[1] College of Telecommunications and Information Engineering, Nanjing University of Posts and Telecommunications, Nanjing, 210003, China
[2] School of Environmental and Municipal Engineering, Xi'an University of Architecture and Technology, Xi'an, 710055, China
[3] Department of Applied Physics and Electronics, Umeå University, 90187 Umeå, Sweden
[4] School of Electrical Engineering and Computer Science, Royal Institute of Technology (KTH), Stockholm, 10044, Sweden
[5] Computer Vision Laboratory, Swiss Federal Institute of Technology (ETH), Zürich, 8092, Switzerland

Corresponding author: Bin Yang (yangbin@xauat.edu.cn, bin.yang@umu.se)

This study was supported by the National Natural Science Foundation of China (No. 61401236), the Jiangsu Postdoctoral Science Foundation (No. 1601039B), the Key Research Project of Jiangsu Science and Technology Department (No. BE2016001-3).

**Abstract:** Human thermal comfort measurement plays a critical role in giving feedback signals for building energy efficiency. A non-invasive measuring method based on subtleness magnification and deep learning (NIDL) was designed to achieve a comfortable, energy efficient built environment. The method relies on skin feature data, e.g., subtle motion and texture variation, and a 315-layer deep neural network for constructing the relationship between skin features and skin temperature. A physiological experiment was conducted for collecting feature data (1.44 million) and algorithm validation. The non-invasive measurement algorithm based on a partly-personalized saturation temperature model (NIPST) was used for algorithm performance comparisons. The results show that the mean error and median error of the NIDL are 0.4834 ºC and 0.3464 ºC which is equivalent to accuracy improvements of 16.28% and 4.28%, respectively.
**Keywords:** Non-invasive method, Thermal comfort perception, vision-based subtleness magnification, Deep learning, Energy efficiency

## 1. Introduction

Real-time thermal comfort perception for occupants plays important roles in human-oriented smart buildings and their energy efficiency. 21% of the global energy consumption is due to energy requirements of commercial and residential buildings [1]. In many countries and regions with rapid urbanization, building energy consumption is expected to increase at an annual rate of 32% [1]. 50% of building energy consumption is related to heating, ventilation and air conditioning (HVAC) system [2]. Feedback signals from thermal comfort perception can be used to effectively control and optimize HVAC energy consumption. Since the 1970s, many methods, including questionnaire surveys, environmental measurements, and physiological measurements (invasive and semi-invasive methods), have been explored to measure human thermal comfort. However, due to (1) inter-individual differences, (2) intra-individual variances, and (3) subtle skin variations (that make it difficult to access skin temperature through computer vision), there has been no breakthrough in thermal comfort perception through computer vision-based techniques until now. The drawbacks of current methods can be summarized as follows: (1) lack of big data validation, (2) lack of practical application possibilities for accurate non-invasive techniques, and (3) lack of adequate consideration of inter-individual and intra-individual differences over time, including subtle skin variations. Instead of human oriented design considering individual perception of indoor climate, buildings are regulated to provide constant and standardized climate comfort. Because different occupants have different subjective feelings toward the same indoor environment, the constant indoor environment parameters cannot meet individual needs in a smart building and optimize energy efficiency.

Human thermal comfort is a subjective feeling that depends on how the human body interacts with the environment [3]. For overcoming the drawbacks described above, a non-invasive measurement method of thermal comfort based on subtleness magnification and deep learning (NIDL) was explored and is described in this paper. The subtleness magnification algorithm adopted is Euler Video Magnification (EVM). Using this NILD method, subtle skin variation was first magnified by the EVM algorithm, and a region of interest (ROI) is selected. A deep neural network with 315 layers was optimized and used for extracting skin image features, according to features of human thermal comfort, and a regression relationship between skin image and skin temperature was constructed. A dataset, containing 1.44 million frames, was collected from a physiological experiment and was used for algorithm validation.

The main contribution of this paper can be summarized as follows:
(1) The proposed method makes non-invasive measuring technology for human thermal comfort practically possible.
(2) It is the first time that deep learning is used for skin temperature measurement using EVM combined with deep learning for feature extraction and relationship construction.
(3) It is the first time that a large image-based dataset (1.44 million frames) for human thermal comfort is

constructed. 16 subjects were invited for a physiological experiment to collect the data set and it was used for algorithm validation.

The rest of this paper is organized as follows. Section 2 introduces related work. In section 3, the research method, including the physiological experiment to collect image-based data and algorithm, are introduced. Algorithm validation results are presented in sections 4 and 5, and conclusions in section 6.

## 2. Related work

Based on Fanger's theory of thermal comfort [3], the thermal comfort environment is defined by ASHRAE and ISO (No. 7730) as "at least 80% of building occupants are psychologically satisfied with the temperature range of thermal environment [4, 5]". As mentioned earlier, human perception thermal comfort varies intra-individually as well as inter-individually. Tracking these variances has traditionally involved three types of methods (2.1-2.3 below).

### 2.1 Questionnaire method

Based on an offline or online questionnaire, the thermal preference of an occupant is collected and used as a basis for environment parameter regulation [6, 7]. According to [4] and [5], the questionnaire is a subjective assessment which can reflect the occupant's psychological state and thermal comfort level. However, the questionnaire method relies on the continuous and frequent participation of building occupants, therefore the operability is weak and the efficiency is low [8].

### 2.2 Environmental measurement method

For the environmental measurement method, different sensors measure different indoor environmental parameters, including temperature, humidity, and airflow. Based on supervision models, the relationship between the environmental parameters and occupant thermal comfort is constructed to determine the comfort level of the building environment. Liu [9] divided the indoor environment into three levels, which are comfortable, uncomfortable warmth, and uncomfortable coolness, and conducted subjective experiments. The statistics data of subjective feeling was collected through subjective voting. Based on it, a neural network with 5 hidden layers was trained with three types of data: air temperature, radiation temperature and air flow. However, intra-individual human thermal comfort variations were not considered. Based on the environmental measurement method, ASHRAE standard 55 [4] and 62.1 [10] defines an indoor environment with constant parameters, which is commonly used by the building industry. The constant parameters include temperature, humidity, and airflow rate. For example, the indoor temperature range in Swedish buildings is controlled between 24 ºC and 27 ºC (collected in KTH campus, Stockholm, Sweden). The winter heating temperature is required from 16 °C to 24 °C in China [11]. In practical operation, the indoor temperature sometimes reaches 27 °C or even 30 °C [12]. Intra- and inter-individual differences and occupants are not considered [8]. Data shows that even a slight indoor temperature adjustment (e. g., 1 °C) has a large impact on the energy consumption of buildings [13, 14]. Therefore, if heating or cooling can be implemented according to individual requirements for thermal comfort, the energy distribution could be better optimized. Farhan [15] uses the RP-884 database to classify individual levels of thermal comfort with support vector machines (SVM). Based on the indoor environment parameters, including air temperature, mean radiant temperature, relative humidity, air velocity and metabolism, Megri [16] estimated some thermal comfort indices which are predicted mean votes (PMV) and predicted percentage dissatisfied (PPD). An optimized SVM and nonlinear kernel function were used. Peng [17] used SVM to construct individual's thermal comfort model. SVM and linear discriminant analysis (LDA) were combined to improve the model efficiency and classify thermal comfort. Peng [18] proposed a demand-driven control strategy for energy saving of HVAC system. Unsupervised and supervised learning were used for model training and the data captured from single-person offices, multi-person offices, and meeting rooms.

### 2.3 Physiological measurement method

Various sensors are used for collecting human physiological parameters such as skin temperature and pulse rate. Such physiological measurements complement subjective instruments such as questionnaires and environmental measures.

**(1) Invasive measuring method**

Wang [19] measured the skin temperature of different parts of the human body. It was found that the correlation is strong between the temperature of finger, the temperature gradient of the fingertip and human thermal comfort. The corresponding correlation coefficients were 0.78 and 0.8, respectively. Yao [20] presented an adaptive measurement model based on PMV. Using the PMV value as a priori knowledge, a fitness coefficient can be calculated with the aid of climate and other parameters. Yao [21] also studied EEG data and Heart Rate Variability (HRV) to verify the possibility of characterizing human thermal comfort. The results showed that there was a close relationship between HRV, EEG, and human thermal comfort. HRV was found to be especially relevant to thermal comfort data. Nakayama [22] constructed a relationship between local skin temperature and human thermal comfort, with a mean square error (MSE) less than one. Simone [23] studied a measurement method for thermal comfort based on energy estimation. In [23], individual energy consumption rates were associated with thermal comfort. Considering the convection and radiation exchange between human and environment, one result was that the individual rate of energy consumption increases when the indoor temperature is more than 24 °C or less than 22 °C. Finally, a second-order polynomial model between human comfort and body energy consumption was constructed. Liu [24] studied the possibility of characterizing human thermal comfort with average skin temperature. The skin temperature values were captured from 26 sampling points of the human body and various mathematical combinations were explored, then the mean value of 10 sampling points was selected as the most accurate metric. Bermejo [25] measured occupant thermal sensation through individual behavior in a constant temperature environment, and a thermal comfort estimation algorithm based on adaptive fuzzy logic was constructed. Kingma [26] presented a mathematical model based on heat-sensing neurophysiology, the data of 12 subjects and 8 subjects were defined as training and test set, respectively. The parameters were skin temperature and core temperature. The results show that the mean error is 0.89 and least square error 0.38. Takada [27] defined the average skin temperature and time difference (intra-individual) as parameters, and then predicted transient thermal sensations of occupants. Based on it, a multivariate regression model was constructed. In this method, when the correlation coefficient reaches 0.839, the thermal sensation predicted is considered as strongly correlated. Sim [28] measured skin temperature through wristbands, and 8 subjects were invited to participate in an experiment with different thermal conditions. Based on it, the thermal comfort model was constructed with average skin temperature, temperature gradient, and temperature time difference. Based on a data-driven method, Chaudhuri [29] predict three types of thermal sensation: uncomfortable coolness, comfortable and uncomfortable warmth. The model was constructed with two types of input parameters, which are environment parameters and human thermal sensation. The results were compared with Support Vector Machines (SVM), neural networks and Linear Discriminant Analysis (LDA). The results show that the prediction accuracy is between 73.14% and 81.2%. Dai [30] predicted thermal comfort demands of individual occupants through SVM analysis. The skin temperature values of different points of body were used for model training. Then the individual thermal sensations were classified. Kim [31] proposed a machine-learning based approach to predict individuals' thermal preference. The data was captured from chairs and six different machine learning algorithms were deployed for improving the prediction accuracy.

**(2) Semi-invasive measuring method**

Ghahramani [32] measured human thermal comfort with an infrared thermography sensor, mounted on eyeglasses. The skin temperature was estimated through three sampling points on the occupant's face. In addition, two ways were defined to describe the hot neutral region and to estimate the occupant's thermal comfort at 95% confidence level. Using the raw data in [32], a measuring method for face thermal comfort, based on Hidden Markov Model (HMM), was constructed [33]. Further, three thermal comfort statuses, uncomfortable warmth and comfortable and uncomfortable coolness, were chosen. The method [33] was validated against 10 subjects with an accuracy of 82.8%.

**(3) Non-invasive measuring method**

Cheng and Yang [34] first presented a non-invasive measuring method with mobile phone and computer cameras. It was the first time that Eulerian Video Magnification (EVM) was combined with thermal comfort measurement. Based on this method, a partly-personalized saturation-temperature model (NIPST, $T_i=96.5 \times S_i+b_i$) was developed to predict skin temperatures for young east Asia females. The mean error and maximum error were 0.5774 °C and 3.0748 °C, respectively. In [34], computer vision and building physics were combined to achieve a novel, non-invasive measurement of human thermal comfort in smart buildings.

The advantages and disadvantages of the above methods can be summarized as follows.
1) The questionnaire method can reflect the psychological state of the occupant well. However, more or less continuous feedback from the occupant is required, so the operability is typically weak.
2) The operability of the environmental measurement method is good. Based on it, the indoor environment can be adjusted by parameters such as measured indoor temperature and humidity. However, the individual occupant's subjective experience is not considered in this method.
3) The thermal sensation of the occupant can be reflected well by physiological measurements through body sensors. However, wearing sensors can be invasive to varying degrees, and typically the operability is therefore weak.

Fortunately, non-invasive methods of physiological measurement can register an individual's key thermal data over a distance, without installing sensors on the human body. Computer vision technology (such as vision-based subtleness magnification technology [35-37]) and machine learning (such as deep learning [38-39]) can support such non-invasive measuring of thermal data. The method proposed here is of this last, non-invasive kind.

## 3. Research method

### 3.1. *Physiological stimulus experiment*

In this paper skin temperature is used for characterizing thermal comfort. While thermal comfort is a subjective feeling, skin temperature can be utilized to reflect approximate human comfort level. 16 subjects were invited for participating in a physiological experiment wherein a total of 1.44 million skin images were collected and temperature coded for algorithm validation.

The collection was handled in a special chamber with constant air temperature and humidity (HOBO, model U12-012). The indoor environment parameters, dry-bulb air temperature, and humidity, were 22.2±0.2 ºC and 36.9±2.5%, respectively. The subjects had the following characteristics: (1) they were Asian females. (2) Their ages were 23.9±3.9 years. (3) Their height were 1.62±0.05 m. (4) Their weight: 52.2±6.5 kg. (5) Their body mass index (BMI): 19.9 ± 2.2 kg/m$^2$. A normal cell phone camera (Huawei, 1280 × 720) was used for collecting skin variation video. The real skin temperature were captured from the back of subjects' hands, with an iButton sensor (model DS192H, error is ±0.125 ºC).

The details of the physiological experiment are summarized as follows: **(1) Experimental preparation**. The environment parameters of the chamber were measured and fine-tuned to ensure the stability of the indoor environment. **(2) Chamber adaptation.** When the subjects arrived at the experiment chamber, they rested for 10 min before the experiment. **(3) Thermal stimulus**. Both hands of the subjects were immersed into 45 ºC warm water for 10 min. **(4) Data collection.** Two kinds of data were captured, including skin variation video and skin temperature. For each subject, 50 min video of skin variation was collected. The corresponding frame rate is 30 frames/second, so that the image data of each subject is 90,000 frames. In this paper, we assume that the skin variations in both hands are the same. The skin variation video was captured from the left hand, and the skin temperature was collected from the right hand. The sampling frequency of iButton was 1 time/min.

### 3.2. *NIDL Algorithm*

When a human thermal or cold reaction occurs, there will be a variation in the skin, such as pore shrinkage, color variation, etc. We deployed Euler Video Magnification (EVM) for amplifying the subjects' skin variations. Let $f(x, t)$ denote human skin image, and $k(t)$ denote the skin variation in different environments. We then have [35, 37]

$$f(x,t) = Z(x + k(t)) \tag{1}$$

where $Z$ denotes the relationship function between $f(x, t)$ and skin variation $k(t)$. Let $\beta$ denote the magnification coefficient, and after first-order Taylor series expansion to Z, we have [35, 37]

$$\begin{aligned}f(x,t) &\approx Z(x) + (1+\beta)k(t)\frac{\partial f(x)}{\partial x} \\ &= Z(x + (1+\beta)k(t))\end{aligned} \tag{2}$$

where, $k(t)$ is the space variation of skin texture at time $t$ and it will be amplified by EVM to a magnitude of $1+\beta$. Equation (2) shows that the invisible skin variation can be amplified to be visible. The collected video data was processed by EVM and considered as the input signal for the deep neural networks.

In this paper, deep learning was used for big data training and the model generated. The Inception deep neural network architecture [39] was adopted and optimized. As shown in Fig. 2, the main function of Inception was retained but the last fully connected layer was removed. Then, an average pooling layer and three fully connected layers were added. The corresponding activation function was a Rectified Linear Unit (ReLU). In order to obtain better prediction results, based on piecewise stationary theory, a calibration function was deployed for the skin temperature predicted by the deep neural network. The calibration function is shown below.

$$T_p' = T_p - \xi \tag{3}$$

where, $T_p$ is the skin temperature predicted by the deep neural network, $T_p'$ is the calibration value of skin temperature. $\xi$ is the calibration coefficient obtained from (4).

$$\xi = \begin{cases} \dfrac{Error_{train\_mean}}{\tau_1}, & if\ \dfrac{1}{n}\sum_{i=1}^{n} Error_{test}(i) > \eta \\ \dfrac{Error_{train\_mean}}{\tau_2}, & if\ Error_{train\_mean} >= \varepsilon \end{cases} \tag{4}$$

where, $Error_{train\_mean}$ is the mean error for the training set, $Error_{test}$ is the error for the test set. The mean of the first $n$ error of test set is considered as prior knowledge and is used to determine the condition of the calibration. The parameters, $\eta$ and $\varepsilon$, are controlled threshold values. The parameters, $\tau_1$ and $\tau_2$, are accuracy adjustment factors. All the parameters, including $\eta, \varepsilon, \tau_1, \tau_2$ and $n$, are defined based on data training and model generation.

In (4), the *Error* is absolute error and used for assessing algorithm performance, as shown in (5)

$$Error(i) = |T(i) - T(i)_r| \quad i = 1, 2, 3, \ldots, t \tag{5}$$

where, $T(i)$ is the skin temperature predicted by the NIDL presented in this paper, $T_r(i)$ is the skin temperature captured by the iButton sensor which is considered ground truth in this paper. $t$ denotes the sampling time. The steps of the NIDL algorithm are summarized in Table 1.

## 4. Results

To validate the NIDL proposed in this paper, a big dataset (1.44 million frames) was collected and used for algorithm training and validation. A total of 16 female subjects participated in the physiological data collection and the skin variation and temperature data was captured from the back of the hands.

The hardware, used for running core code and training data, was an X64-based workstation with 32G RAM, double Processors. and Graphics Processing Unit (GPU). The processors are Intel (R) Xeon (R) CPU E5-2687W V3 @ 3.10GHz and the GPU is NVIDIA GeForce GTX 980 (1920 × 1080, 32 bit, 60Hz).

As shown in Fig. 3, the hand/skin images (1.44 million images from 16 subjects) were amplified through Euler Video Magnification (EVM). '1 Raw' denotes the raw hand images, '1 EVM' denotes the magnified hand images and '1' denotes the subject number. The hand images with thermal comfort information were processed by EVM, so that the texture features of human skin variations were suitably amplified. The data, magnified hand images and skin temperatures captured by iButton sensor, were combined together in a labeled document. Then the data was fed to a deep neural network with 315 hidden layers for model training.

To achieve improved model performance, the transfer learning network, Inception V3, was optimized. Let $n$ denote batch size, and each time $n$ images will be imported into the workstation stack. For improving the prediction accuracy, we removed the last fully connected layer of Inception V3. If the size of input data is $n \times 150 \times 150 \times 3$, then the output data size of Inception will be $n \times 1 \times 1 \times 2048$. Based on this, we added a pooling layer, and the average pooling algorithm was adopted. As such, the corresponding output is $n \times 2048$. Further to this, three fully connected layers were stacked layer by layer to place on top of the average pooling layer, with sizes 2048 × 1024, 1024 × 512 and 512 × 1, respectively. The corresponding activation function is a Rectified Linear Units (ReLU). As a result, the output of the three Fully Connected Layers are then $n \times 1024$, $n \times 512$ and $n \times 1$, respectively. To prevent stack overflow, $n$ is defined as 50 in this paper. This means that 50 images, taken from the training set, were imported into the workstation stack in each loop. The corresponding output of the whole deep neural network is the prediction value of skin temperature of the images fed to the network, and the size of output vector is $50 \times 1$.

For every 100 loops, in which 5000 images were trained, one time performance verification was handled. Then, fixed 50 images from testing set were used for absolute error computing. If the mean error is less than $\varepsilon$, then we will save the current algorithm model, or we will save the algorithm model when one 'epoch' is finished. In addition, when loops are 100000 or 20000, we will save the algorithm model as well. It should be noted that one 'epoch' means that the whole data of training set is trained for one time. In this paper, $\varepsilon$ is defined as 0.3 ℃ and epoch is 7.

For algorithm performance comparison, the non-invasive measuring method based on a partly personal ST model (NIPST) was used in this paper. Fig. 4 shows the results of subjects No. 1-8 and Fig. 5, the results of subjects No. 9-16. The prediction values for skin temperature are shown in the left column (Fig. 4-a, c, e, g, I, k, m, o, Fig. 5-a, c, e, g, I, k, m, o) and the corresponding errors are shown in the right column (Fig. 4-b, d, f, h, j, l, n, p, Fig. 5-b, d, f, h, j, l, n, p). It was shown that the prediction skin temperatures float up and down around the real skin temperatures captured by the iButton sensor. For the 16 subjects, the mean errors in ℃ were 0.5100, 0.3272, 0.2501, 0.3103, 0.2964, 0.2012, 0.2383, 0.2682, 0.4061, 0.3414, 0.3497, 0.5022, 0.7275, 0.8089, 1.8983, and 1.7733. 10 of 16 subjects' errors were less than 0.5 ℃ and 2 of 16 subjects' errors were approximately equal to 0.5 ℃.

Table 1 shows the error mean, median and standard deviation of the NIPST algorithm as 0.5774 ℃, 0.3619 ℃, and 0.6118 ℃, respectively. Based on the NIDL algorithm presented in this paper, the three parameters are 0.4834 ℃, 0.3464 ℃ and 0.4959 ℃, respectively. Although the error maximum of NIDL is bigger than that of NISPT, the first few large errors of NIDL fall quickly. The 3$^{rd}$, 4$^{th}$, 5$^{th}$ largest errors of NIDL are 3.0108 ℃, 2.5974 ℃ and 2.4793 ℃, respectively. However, the 3$^{rd}$, 4$^{th}$, 5$^{th}$ largest errors of NIPST are 3.0298 ℃, 2.9525 ℃ and 2.9278 ℃, respectively.

The error distribution comparison was shown in Fig. 6 and Table 3. As to the NIPST algorithm, the errors less than 0.3 ℃ account for 44.7122% of all the data, however, the same error ratio of NIDL account for 45.3846% of all the data. The error between 0.3 ℃ and 0.5 ℃, 17.57% is improved by the NIDL model. The error between 0.5 ℃ and 1.0 ℃, 8.46% is also improved by NIDL model. For all the errors more than 1 ℃, the proportion of the NIDL presented is significantly reduced. Four error intervals ([1.0 1.5], [1.5 2.0], [2.0 2.5], [2.5 4.2]) are reduced respectively 4.23%, 48.23%, 60.57% and 64.09%. Fig. 6 and Table 3 show that the error of NIDL presented is mainly less than 1 ℃ and the performance of NIDL is better than that of NIPST.

**TABLE 1.** Non-invasive thermal comfort perception based on subtleness magnification and deep learning

| Algorithm: the NIDL algorithm |
|---|
| **Output:** NIDL model (*.h5*), skin temperature (°C) |
| **Step:** |
| 1. Video data preprocessing and texture magnification |
|    (1) De-noise for skin variation video. |
|    (2) Texture magnification for 1.44 million video data (frames), $\beta = 10$ (equation 2). |
| 2. ROI extraction |
|    (1) Frame extraction: hand images were extracted from video after texture magnification. |
|    (2) ROI extraction: local area of the back of hand images (150×150) is considered as ROI. |
| 3. Temperature interpolation: linear interpolation was used for skin temperature captured by the iButton sensor. |
| 4. Algorithm training with big dataset and deep learning |
|    (1) Labeling: making label document for ROI images of extraction and interpolated temperature. |
|    (2) Testing set: the data of 1 subject (for each round a different subject was chosen). |
|    (3) Training set: the data of the remaining 15 subjects. |
|    (4) For $j = 1$: *epoch* (*epoch* = 7) |
|        For $m = 1$: *loops* (*loops* = 1.44 million/50) |
|        1) Input 50 ROI images of extraction into an optimized Inception network. |
|        2) If *m* is an integer multiple of 100 (importing 5,000 images), the performance verification will be handled. Then, a set of 50 images from the testing set were used for performance verification during model training. |
|        3) If *Error* $< \varepsilon$ ($\varepsilon = 0.3$ °C), then the current model is saved. |
|        4) Or if *m* is an integer multiple of 10000 then the current model is saved. |
|        5) Or if all the training set is used for training in the $j^{th}$ epoch (*m* is equal to *loops*) then the current model is saved. |
|        End |
|      End |
| 5. Optimizing model parameters and calibrating for skin temperature prediction. The parameter in formula (4) are shown as follows (1) $\tau_1 = 0.446$ (2) $\tau_2 = 12.88$ (3) $n = 3$ (4) $\eta = 1$ °C, and the $\varepsilon$ is the same as step 4 which is 0.3 °C. |

**TABLE 2.** Error comparison (maximum, minimum, mean, median and standard deviation)

|  | NIPST (°C, Baseline) | NIDL (°C, This paper) | parameters variation | Performance Optimization or not |
|---|---|---|---|---|
| Maximum | 3.0748 | 4.1417 | / |  |
| 2nd largest | 3.0404 | 3.3912 | / |  |
| 3rd largest | 3.0298 | 3.0108 | / |  |
| 4th largest | 2.9525 | 2.5974 | / |  |
| 5th largest | 2.9278 | 2.4793 | / |  |
| Minimum | 0 | 1.0e-4 | / |  |
| Mean | 0.5774 | 0.4834 | ↓ 16.28% | ↑ |
| Median | 0.3619 | 0.3464 | ↓ 4.28% | ↑ |
| Standard deviation | 0.6118 | 0.4959 | ↓ 18.94% | ↑ |

**TABLE 3.** Absolute error distribution

| Absolute error (°C) | NIPST (Baseline, %) | NIDL (This paper, %) | Distribution Variation | Performance Optimization or not |
|---|---|---|---|---|
| [0, 0.3) | 44.7122 ⎫ | 45.3846 ⎫ | ↑1.5% | ↑ |
| [0.3, 0.5) | 16.4659 ⎬ 61.1780 | 19.3590 ⎬ 64.7436 | ↑17.57% | ↑ |
| [0.5, 1.0) | 22.2222 | 24.1026 | ↑8.46% | ↑ |
| [1.0, 1.5) | 7.2289 | 6.9231 | ↓4.23% | ↑ |
| [1.5, 2.0) | 4.9531 | 2.5641 | ↓48.23% | ↑ |
| [2.0, 2.5) | 2.2758 | 0.8974 | ↓60.57% | ↑ |
| [2.5, 4.2) | 2.1419 | 0.7692 | ↓64.09% | ↑ |

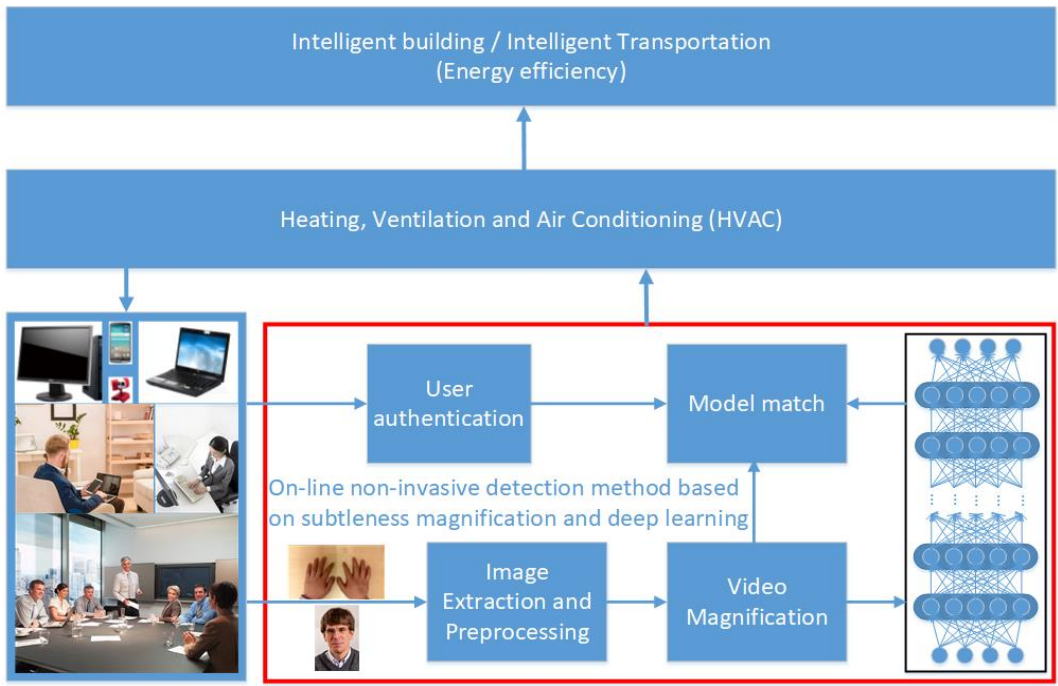

Fig. 1. Schematic of non-invasive perception of thermal comfort in practical application.

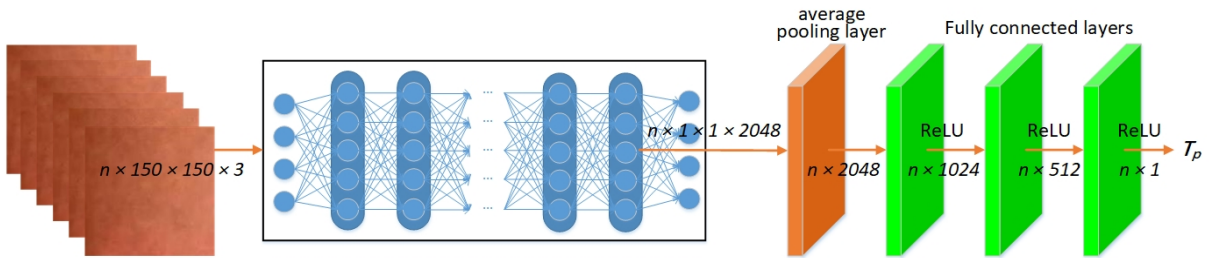

Fig. 2. The NIDL network framework with optimized Inception

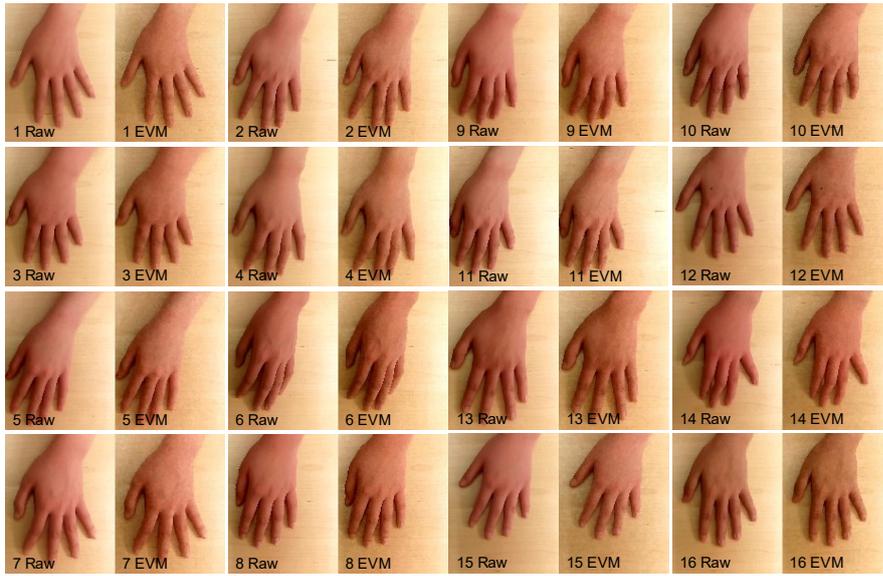

Fig. 3. The backs of 16 hands (raw hand images and magnified hand images by EVM)

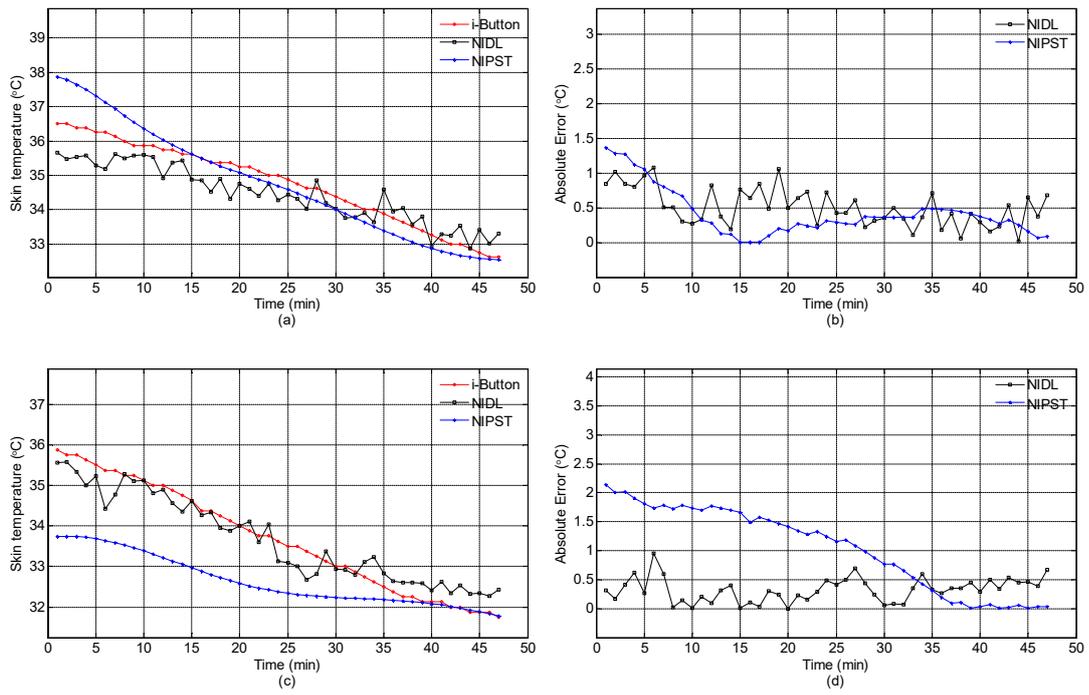

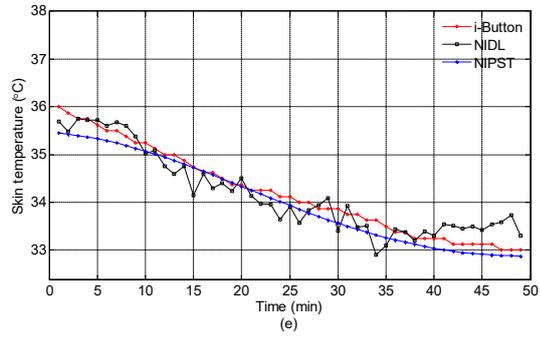
(e)

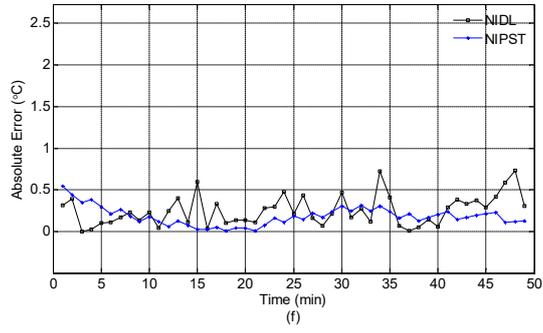
(f)

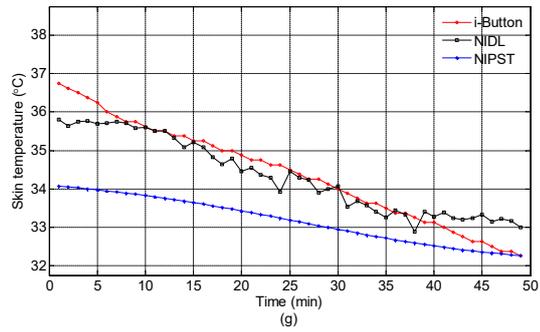
(g)

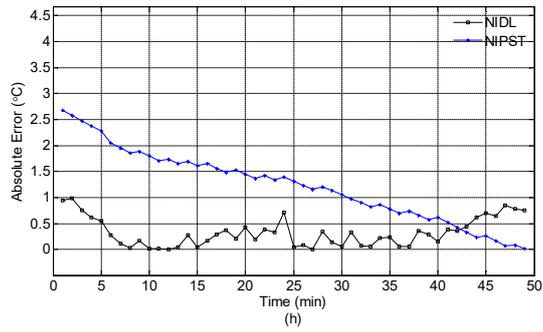
(h)

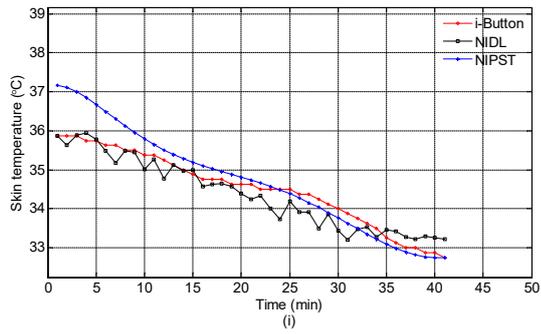
(i)

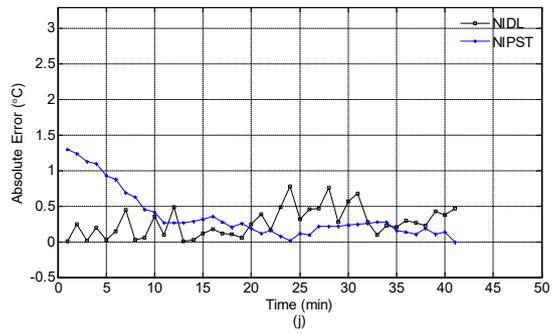
(j)

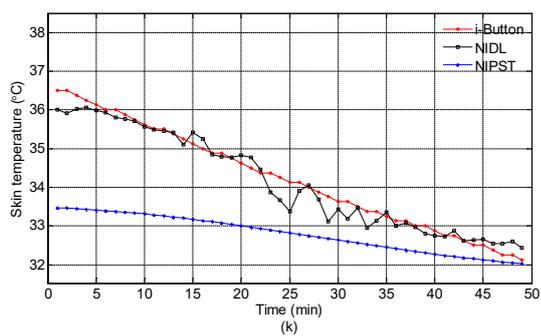
(k)

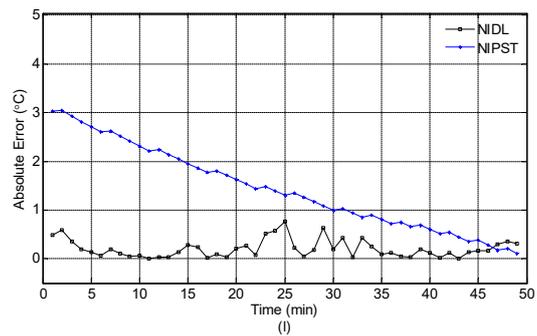
(l)

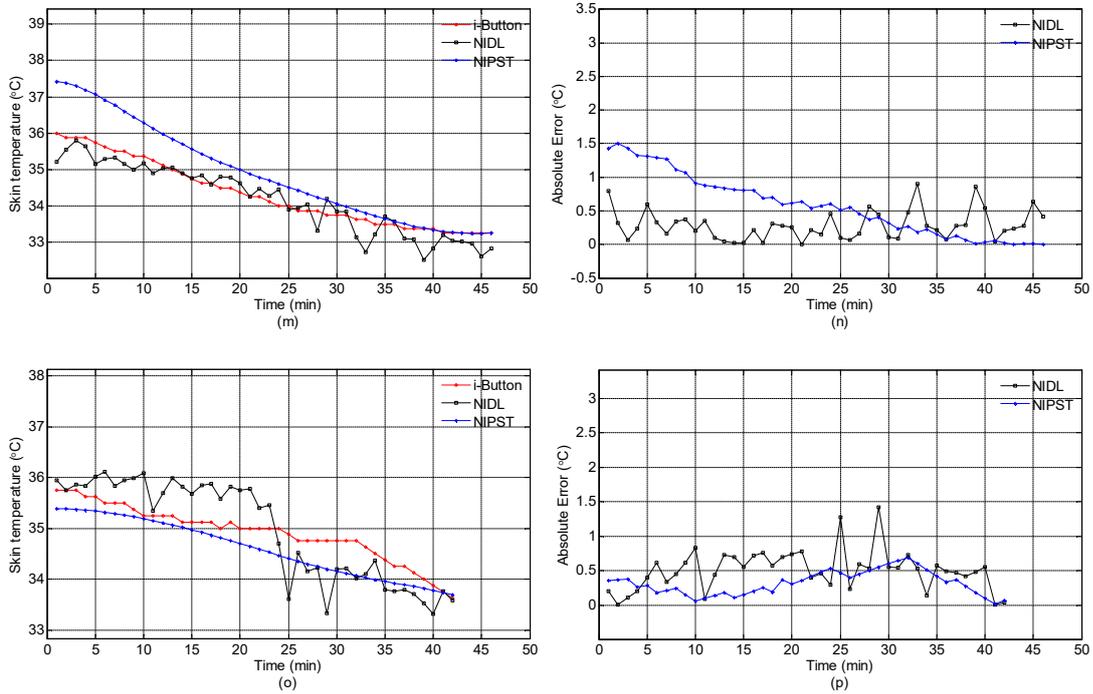

Fig. 4. Validation results comparison between NIDL, NIPST and the iButton sensor (The data of subjects No. 1-8 are shown in Fig. 4. A total of 16 subjects' hands were stimulated with hot water. Cross validation was adopted and 16 rounds of model training were carried out. For each round of training, one subject's data was defined as the test set, and the remaining 15 subjects' data was defined as the training set. The charts on the left, Fig. 4-a, c, e, g, i, k, m, o, show the variation curves of skin temperature, while the charts on the right, Fig. 4-b, d, f, h, j, l, n, p, show the absolute errors.)

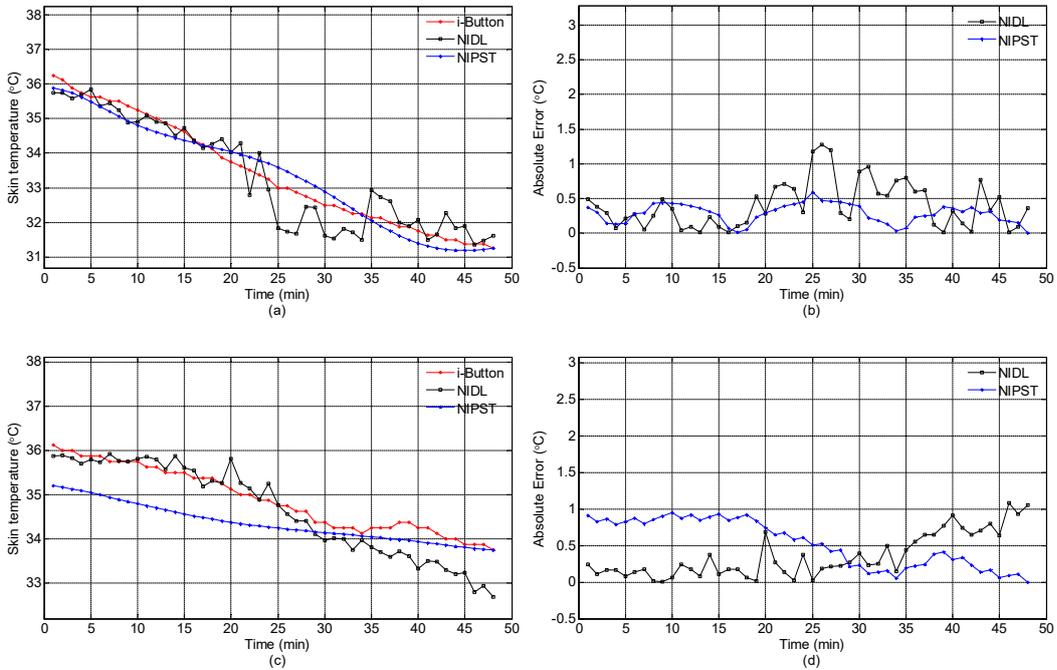

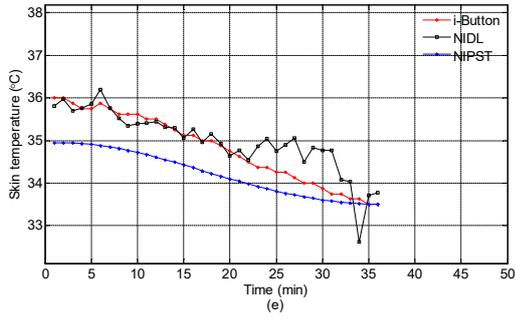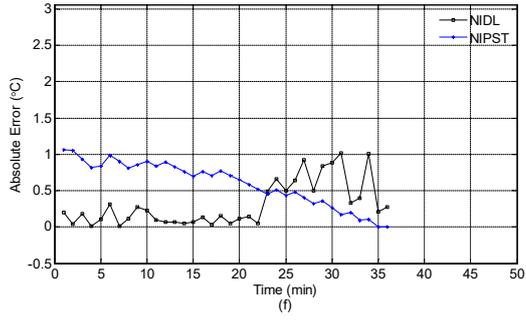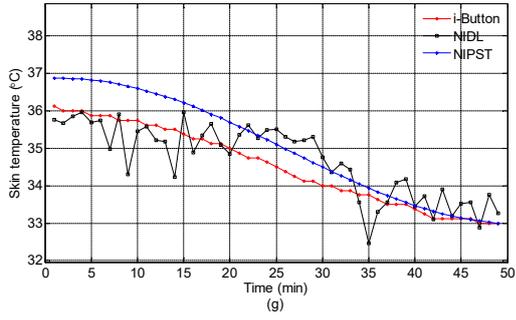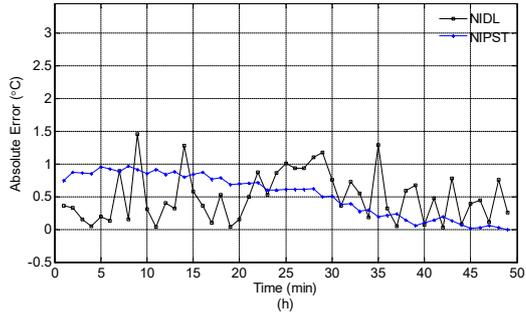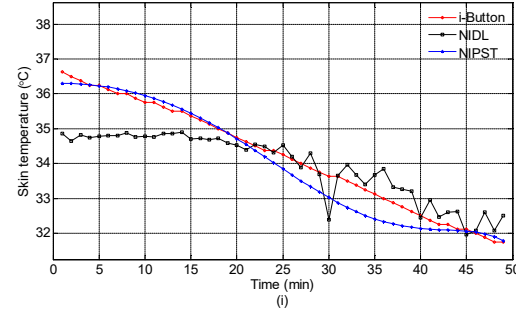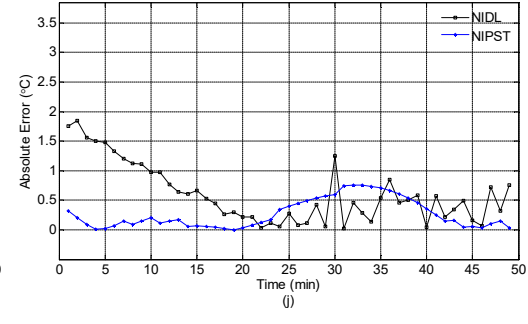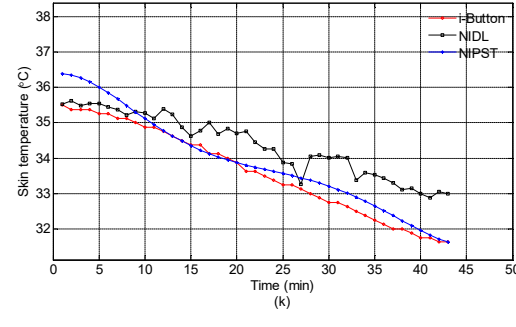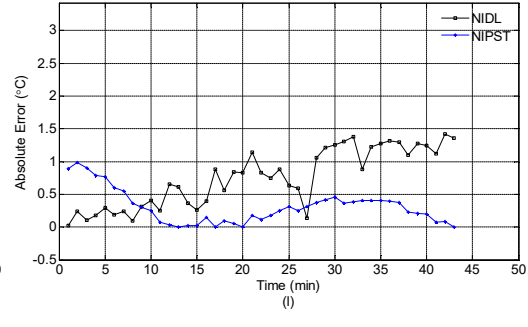

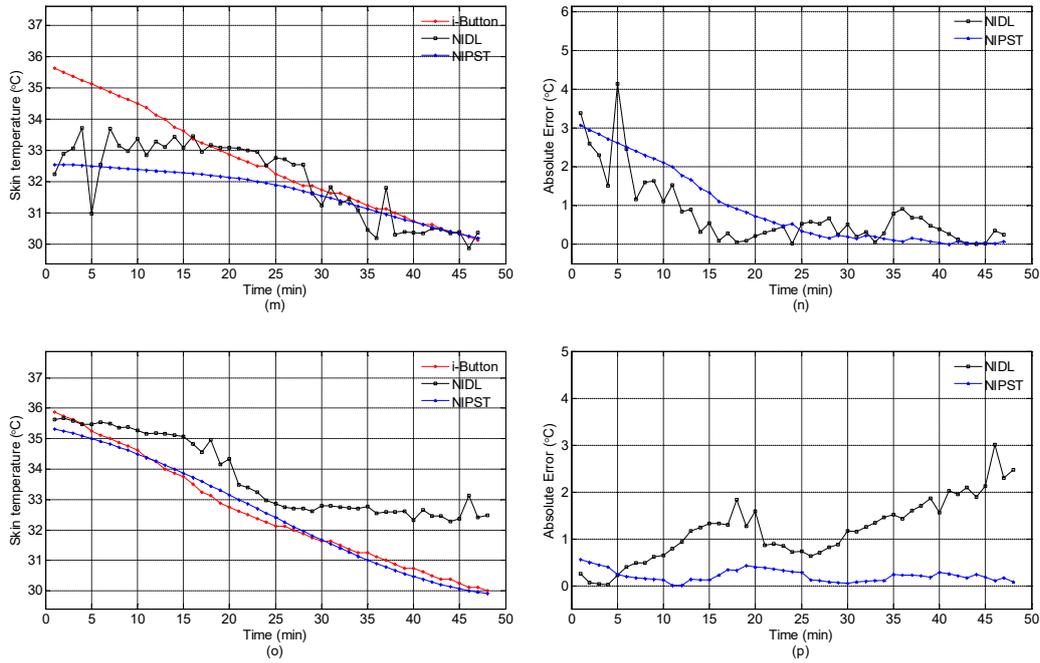

Fig. 5. Validation results comparison between NIDL, NIPST and the iButton sensor (The data of subjects No. 9-16 are shown in Fig. 5. A total of 16 subjects' hands were stimulated with hot water. Cross validation was adopted and 16 rounds of model training were carried out. For each round of training, one subjects' data was defined as the test set, and the remaining 15 subjects' data was defined as the training set. The charts on the left, Fig. 5-a, c, e, g, i, k, m, o, show the variation curves of skin temperature, while the charts on the right, Fig. 5-b, d, f, h, j, l, n, p, show the absolute errors.)

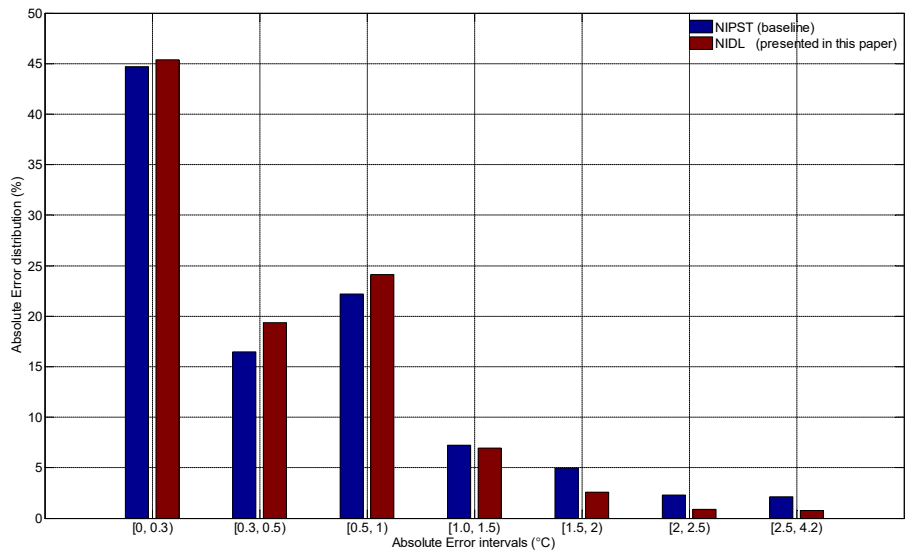

Fig. 6. Absolute error distribution comparison between NIPST and NIDL.

5. Discussion

The NIDL algorithm proposed in this paper has generalizability because the algorithm model generated by the data of 15 subjects can be applied to the skin temperature prediction of another subject. Further, cross-validation is adopted in this paper and 16 rounds of model training experiments are handled. In each round, the video data, 1 of 16 subjects, was defined as testing set, respectively. Then, the remaining data of 15 subjects were used as training set. In addition, mean errors of each testing set were calculated. 14 mean errors were less than 1 ℃. Further, 10 of 14 mean errors were less than 0.5 ℃ and 2 of 14 mean errors were approximately equal to 0.5 ℃. The results in Fig. 4, 5, 6 show that the performance of NIDL was better that of NIPST. Based on the mean error and median error of NIDL (0.4834 ℃, 0.3464 ℃) shown in Table 2, it is clear that NIDL is promising with respect to practical application.

The main difference between the NIPST model [34] and the NIDL model presented in this paper can be summarized as follows (1) **Model type**: the NIPST is a linear model ($T_i = 96.5 \times S_i + b_i$) and polynomial regression was used for generating it. The NIDL is a nonlinear model generated by an optimized deep neural network with 315-layers; (2) **Generalizability**: the NIPST model does not have generalizability. Because the parameters are constant and limited, the error will increase if the NIPST model generated is used for predicting skin temperature of another occupant. Fortunately, cross-validation showed that The NIDL has this kind of generalizability. (3) **Application convenience:** for the NIPST model, many images and corresponding skin temperature data is required to calculate parameters of individual occupants. However, it is impossible to capture real temperature for every occupant in practical application. For the NIDL model, the input data are skin images, and the predication values of skin temperature are then derived from then. This process lends itself well for practical application. Based on the comparison above, deep neural network and big data training also improve the adaptability and applicability of the NIDL proposed.

How can the challenges associated with inter-individual and intra-individual variations be overcome? Firstly, the EVM algorithm was used in NIDL to magnify the skin texture, so that the challenge of 'subtle skin variation' is overcome. Secondly, in a practical application, the NIDL will be used online, so that skin images can be captured continuously. Thus the real-time thermal comfort of occupant can be captured and the challenge of time-varying intra-individual differences be overcome. Finally, large datasets contains a variety of thermal comfort features. In a practical application, more data could be used for model training and more kinds of intra- and inter-individual features be extracted into the NIDL model, making it more accurate.

Some researchers may argue that deep learning can achieve skin temperature prediction in this paper directly. Why do we use video magnification technology for amplifying the skin variation? The main reason is that deep learning cannot solve all the problems. Skin variation are very subtle in normal office environments. Therefore, skin textures should be magnified first, and then the features be extracted and analyzed by means of deep neural networks.

Some may say that the research in this paper is merely an application of deep learning. However, this is not the case. Firstly, from the perspective of computer vision, video magnification technology and deep learning are combined to detect thermal comfort. Secondly, the Inception platform was optimized in this paper. Further to this, from the perspective of building physics, the main contribution of this paper is that we explore non-invasive measuring technology for future deployment in a practical application. In fact, apart from one study [34], there is no existing practically useful non-invasive measuring method of thermal comfort. As shown in Table 2, the mean error of NIPST [34] is 0.5774 ℃, however, the mean error of NIDL is only 0.4834 ℃.

It should be noted that in all the 16 mean errors of the testing set described above, there are still 2 mean errors of NILD that are more than 1 ℃. The main reason is the limited volume of data collected. How could deep learning work so well in our research? One important reason is that the volume of big data was sufficient so that the deep neural network could learn more features from the data. In this paper, we collected approximately 1.44 million frames data from physiological experiment. However, the number of subjects is only 16, so the data diversity is still limited. With more data diversity, the approach could work even better.

6. Conclusion

Our aim was to research a novel non-invasive measuring method for thermal comfort. Based on video magnification and transfer learning, the NIDL was presented, and a physiological experiment was conducted, so that the algorithm could be validated. The conclusion can be summarized as follows.

(1) Deep learning can be combined with EVM for extracting skin features and estimating the skin temperature.
(2) The NIDL model generated in this paper can be applied for non-invasive measurements for female Asians.
(3) More data will contribute to generating a better NIDL model.

It should be noted that individual differences are significant in a practical application. People differ in their sensitivity to cold and heat. An individual sensitive index will be produced in our future work to further overcome the challenge of individual differences.

7. Acknowledgement

The authors thanks to William T. Freeman (MIT) for providing his MATLAB code of Euler Video Magnification (EVM). Thanks to Dr. Dengxin Dai (ETH), Dr. Wen Li (ETH), Dr. Ajad Chhatkuli (ETH) and Erik Isaksson (KTH) for their comments.